\documentclass[structabstract]{aa}  
\usepackage{graphicx}
\usepackage{txfonts}
\usepackage{natbib}
\usepackage{amssymb}
\usepackage{amstext}
\usepackage{mathabx}

\usepackage{multirow}

\usepackage{ulem}
\usepackage{lineno}

\newcommand{\unitcm}[1]{cm$^{#1}$}

\newcommand{\unitgamma}[0]{cm$^{-1}\cdot$atm$^{-1}$}

\newcommand{\yco}{$y_{\mathrm{CO}}$}

\newcommand{\Flux}[1]{$\phi_{\mathrm{#1}}$}

\newcommand{\Fluxunit}{\,cm$^{-2}\cdot$s$^{-1}$}
\newcommand{\Kunit}{\,cm$^{2}\cdot$s$^{-1}$}

\newcommand{\dix}[1]{$\times10^{#1}$}
\newcommand{\fig}[1]{Fig.~\ref{#1}}

\newcommand{\tab}[1]{Table~\ref{#1}}

\begin{document}

   \title{First submillimeter observation of CO in the stratosphere of Uranus\thanks{{\it Herschel} is an ESA space observatory with science instruments provided by European-led 
   Principal Investigator consortia and with important participation from NASA.}}    

   \author{T.~Cavali\'e\inst{\ref{inst1},\ref{inst2}}
          \and
          R.~Moreno\inst{\ref{inst3}}
          \and
          E.~Lellouch\inst{\ref{inst3}}
          \and
          P.~Hartogh\inst{\ref{inst4}}
          \and
          O.~Venot\inst{\ref{inst5}}
          \and
          G.~S.~Orton\inst{\ref{inst6}}
          \and
          C.~Jarchow\inst{\ref{inst4}}
          \and
          T.~Encrenaz\inst{\ref{inst3}}
          \and
          F.~Selsis\inst{\ref{inst1},\ref{inst2}}
          \and
          F.~Hersant\inst{\ref{inst1},\ref{inst2}}
          \and
          L.~N.~Fletcher\inst{\ref{inst7}}
          }

   \institute{Univ. Bordeaux, LAB, UMR 5804, F-33270, Floirac, France\\
                   \email{cavalie@obs.u-bordeaux1.fr}\label{inst1}
         \and 
                  CNRS, LAB, UMR 5804, F-33270, Floirac, France\label{inst2}
         \and 
                  LESIA--Observatoire de Paris, CNRS, Universit\'e Paris 06, Universit\'e Paris--Diderot, Meudon, France\label{inst3}
         \and 
                  Max Planck Institut f\"ur Sonnensystemforschung, Katlenburg-Lindau, Germany\label{inst4}
         \and 
                  Instituut voor Sterrenkunde, Katholieke Universiteit Leuven, Leuven, Belgium\label{inst5}
         \and 
                  Jet Propulsion Laboratory, California Institute of Technology, Pasadena, USA\label{inst6}
         \and 
                  Atmospheric, Oceanic and Planetary Physics, Clarendon Laboratory, University of Oxford, Parks Road, Oxford, OX1 3PU, UK\label{inst7}
                 }

   \date{Received 17 July 2013 / Accepted 7 November 2013 }

 
  \abstract
   {Carbon monoxide (CO) has been detected in all Giant Planets and its origin is both internal and external in Jupiter and Neptune. Despite its first detection in Uranus a decade ago, the magnitude of its 
   internal and external sources remains unconstrained.}
   {We targeted CO lines in Uranus in the submillimeter range to constrain its origin.}
   {We recorded disk-averaged spectra of Uranus with a very high spectral resolution at the frequencies of CO rotational lines in the submillimeter range in 2011-2012. We used empirical and diffusion 
   models of the atmosphere of Uranus to constrain the origin of CO. We also used a thermochemical model of its troposphere to derive an upper limit on the O/H ratio in the deep atmosphere of Uranus.}
   {We have detected the CO(8-7) rotational line for the first time, with Herschel-HIFI. Both empirical and diffusion model results show that CO has an external origin. An empirical profile in which CO 
   is constant above the 100\,mbar level with a mole fraction of 7.1-9.0\dix{-9} (depending on the adopted stratospheric thermal structure) reproduces the data. Sporadic and steady source 
   models cannot be differentiated with our data. Taking Teanby \& Irwin's internal source model upper limit of a mole fraction of 2.1\dix{-9} [Teanby \& Irwin 2013. ApJ, 775, L49], the deep O/H 
   ratio of Uranus is lower than 500 times solar according to our thermochemical computations.}
   {Our work shows that the average mole fraction of CO decreases from the stratosphere to the troposphere and thus strongly advocates for an external source of CO in Uranus. Photochemical 
   modeling of oxygen species in the atmosphere of Uranus and more sensitive observations are needed to reveal the nature of the external source.}

  
   \titlerunning{First submillimeter observation of CO in the stratosphere of Uranus}
   
   \maketitle
%

\section{Introduction}

The detection of water vapor (H$_2$O) and carbon dioxide (CO$_2$) in the stratospheres of the Giant Planets and Titan by \citet{Feuchtgruber1997}, \citet{Coustenis1998}, 
\citet{Samuelson1983} and \citet{Burgdorf2006} has raised several questions: what are the sources of oxygen to their upper atmospheres? And do the sources vary from planet to 
planet? Oxygen-rich deep interiors of the Giant Planets cannot explain the observations because these species are trapped by condensation below their tropopause (except CO$_2$ 
in Jupiter and Saturn). Therefore, several sources in their direct or far environment have been proposed: icy rings and/or satellites \citep{Strobel1979}, interplanetary dust particles 
\citep{Prather1978} and large comet impacts \citep{Lellouch1995}.

While the relative similarity of the infall fluxes inferred for H$_2$O by \citet{Feuchtgruber1997} may indicate that interplanetary dust particles (IDP) could be the source for all Giant 
Planets \citep{Landgraf2002}, infrared and far-infrared observations have unveiled a quite different picture. ISO, Cassini, Odin and Herschel observations have proven that the Jovian 
stratospheric H$_2$O and CO$_2$ originate from the Shoemaker-Levy~9 (SL9) comet impacts \citep{Lellouch2002,Lellouch2006,Cavalie2008b,Cavalie2012,Cavalie2013}, while 
Herschel has recently shown that the external flux of water at Saturn and Titan is likely due to the Enceladus geysers and the water torus they feed \citep{Hartogh2011,
Moreno2012}. 

The situation is even more complex for carbon monoxide (CO). Because CO does not condense at the tropopauses of Giant Planets, oxygen-rich interiors are also a potential source. 
An internal component has indeed been observed in the vertical profile of CO in Jupiter by \citet{Bezard2002} and in Neptune, originally by \citet{Marten1993} and \citet{Guilloteau1993}, 
while an upper limit has been set on its magnitude by \citet{Cavalie2009} and \citet{Fletcher2012} for Saturn. The measurement of the tropospheric mole fraction of CO can be used to 
constrain the deep O/H ratio \citep{Lodders1994}, which is believed to be representative of condensation processes of the planetesimals that formed the Giant Planets \citep{Owen1999,
Gautier2005}. On the other hand, large comets seem to be the dominant external source, as shown by various studies: \citet{Bezard2002} and \citet{Moreno2003} for Jupiter, \citet{Cavalie2010} 
for Saturn and \citet{Lellouch2005,Lellouch2010}, \citet{Hesman2007}, \citet{Fletcher2010} and \citet{Luszcz-Cook2013} for Neptune.

The first detection of CO in Uranus was obtained by \citet{Encrenaz2004} from fluorescent emission at 4.7\,$\mu$m. Assuming a uniform distribution throughout the atmosphere, a mixing 
ratio of 2\dix{-8} was derived. The authors tentatively proposed that CO was depleted below the tropopause, suggesting that CO would have an external origin. Despite this first detection 
almost a decade ago, the situation has remained unclear. Ground-based heterodyne spectroscopy has been used unsuccessfully in the past to to try and detect CO in Uranus. 
\citet{Rosenqvist1992} have first set an upper limit of 4\dix{-8} and subsequent attempts to detect CO have failed so far \citep{Marten1993,Cavalie2008a}. In this paper, we present 
observations of CO in Uranus carried out with the Herschel Space Observatory \citep{Pilbratt2010} in 2011-2012, that led to the first detection of CO in Uranus in the submillimeter range. 

In the following sections, we will describe the observations, their modeling and the new constraints on the origin of CO in Uranus and its deep O/H ratio that we have derived.

\section{Observations \label{Observations}}

\begin{table*}
  \caption{Summary of the Herschel-HIFI observations of CO in Uranus. }             
  \label{Obs_list}      
  \begin{center}          
  \begin{tabular}{ccccccc}
    \hline   
    Date              & OD    & Obs. ID           & $\nu$ [GHz]     & $\Delta t$ [h] & $\theta_{\mathrm{HIFI}}$ [\arcsec] & $\theta_{\mathrm{Uranus}}$ [\arcsec] \\
    \hline                    
    2011-07-01  & 779   & 1342223423 & 921.800\,GHz & 1.82               & 23.0                                                        & 3.53 \\
    2012-06-15  & 1128 & 1342247027 & 921.800\,GHz & 2.54               & 23.0                                                        & 3.47 \\
    2012-06-15  & 1128 & 1342247028 & 921.800\,GHz & 2.54               &23.0                                                         & 3.47 \\
    2012-06-15  & 1128 & 1342247029 & 921.800\,GHz & 2.54               &23.0                                                         & 3.47 \\
    \hline  
  \end{tabular}
  \end{center}
  \small{\underline{Note:} OD means operational day, $\nu$ is the CO line center frequency, $\Delta t$ is the total integration time, $\theta_{\mathrm{HIFI}}$ is the Herschel-HIFI telescope 
  beamwidth, and $\theta_{\mathrm{Uranus}}$ is the equatorial diameter of Uranus.}
\end{table*}

We observed the CO(8-7) line at 921.800\,GHz with the HIFI instrument \citep{Degraauw2010} aboard Herschel \citep{Pilbratt2010} on July 1, 2011, as part of the Guaranteed 
Time Key Program ``Water and related chemistry in the solar system'', also known as ``Herschel solar system Observations'' (HssO; \citealt{Hartogh2009}). The CO(8-7) line was 
targeted in Uranus for $\sim$2 hours. The resulting spectrum led us to a tentative detection of CO in Uranus at the level of $\sim$2.5$\sigma$ (on the line peak) at native resolution 
and encouraged us to perform a deeper integration of this line. 

We obtained a $\sim$8-hour integration (split into three observations of equal length) of the same line on June 15, 2012, as part of the Herschel Open Time 2 program 
OT2\_tcavalie\_6. The observations have been performed in double-beam switch mode with the Wide Band Spectrometer (WBS) at a nominal spectral resolution of 1.1\,MHz 
(more details given in \tab{Obs_list}). We have processed the data with HIPE 9 \citep{Ott2010} up to Level 2 for the H/V polarizations and stitched the WBS subbands together. 
The baseline ripple frequencies caused by the strong continuum emission of Uranus have been determined by a normalized periodogram \citep{Lomb1976} and the 3-4 strongest 
sine waves have been removed. Those sine waves are caused by the hot and cold black bodies and by the local oscillator chain of the instrument  and have periods of 90-100\,MHz 
\citep{Roelfsema2012}. The double-sideband response of HIFI has been corrected by assuming a sideband ratio of 1, i.e., a single sideband gain of 0.5 \citep{Roelfsema2012}, and 
identical continuum levels in both sidebands. The uncertainty on the sideband ratio is 12\% (3\% on the single sideband gain), and the continuum levels in the two sidebands should 
be different by less than 1\%, according to our model. Finally, we have coadded the H/V polarizations after weighting them according to their respective noise levels (the V spectra were 
always noisier than the H spectra). We have obtained a clear detection at the level of 7$\sigma$ on the line peak, at a smoothed resolution of 8\,MHz using a gaussian lineshape, 
on the combined 8-hour observation shown in \fig{empirical_models}. Because we have not performed any absolute calibration, we will analyze this line in terms of line-to-continuum ratio ($l/c$). 
The observed continuum levels differ by 6\% in the H and V polarizations, so we have to account for an additional uncertainty of 3\% on the continuum level of our averaged spectrum. 

We note that we also targeted the CO (3-2) and (6-5) lines (at 345.796\,GHz and 691.473\,GHz, respectively) in Uranus using the HARP receiver array and the D-band receiver (respectively) 
of the James Clerk Maxwell Telescope (JCMT) on October 15-16 and November 2, 2009, as part of the M09BI02 project. These observations resulted in the determination of an upper limit of 
6\dix{-8} uniform with altitude up to the CO homopause for the CO mole fraction and will not be discussed further.

\section{Models and results \label{Modeling}}
  \subsection{Radiative transfer model}
  We have performed all spectral line computations with the forward radiative transfer model detailed in \citet{Cavalie2008a,Cavalie2013}, adapted to Uranus. This line-by-line model accounts 
  for the elliptical geometry of the planet and its rapid rotation. Opacity due to H$_2$-He-CH$_4$ collision-induced absorption \citep{Borysow1985,Borysow1988,Borysow1986} was included. 
  \citet{Orton2007} published H$_2$-H$_2$ collision-induced coefficient tables which better reproduce the continuum of Uranus between 700 and 1100\,\unitcm{-1} as observed by Spitzer, 
  but these coefficients are not significantly different in the wavelength range of our observations. We have used the  JPL Molecular Spectroscopy catalog \citet{Pickett1998} and the H$_2$/He 
  pressure-broadening parameters for CO lines from \citet{Sung2004} and \citet{Mantz2005}, i.e. a collisional linewidth at 300\,K of 0.0661\,\unitgamma~for the CO(8-7) line and a temperature 
  dependence exponent of 0.638. We have used the thermal profiles of \citet{Feuchtgruber2013} and \citet{Orton2013a}. They have the same tropopause temperature (53\,K) but the profile of 
  \citet{Feuchtgruber2013} is continuously warmer than the profile of \citet{Orton2013a} in the stratosphere (by 2\,K at 10\,mbar, 5\,K at 1\,mbar, and 11\,K at 0.1\,mbar). We will present 
  results for both thermal profiles hereafter. All synthetic lines have been smoothed to the working resolution of 8\,MHz using a gaussian lineshape.
  
  The CO line is optically thin with $\tau$$=$0.04-0.25 (depending on models) and probes the stratosphere of Uranus between the 0.1 and 5\,mbar levels, allowing us to derive information 
  on the CO abundance. The S/N of the observations results in error bars of 14\%. By adding this uncertainty quadratically with other uncertainty sources (sideband ratio, continuum levels), we 
  end up with an uncertainty of 19\% on the results presented hereafter.
  
  \subsection{Empirical models \label{empirical}}
  We have tested two classes of empirical models: (i) uniform profiles (referred to as ``Uniform'' hereafter), and (ii) uniform profile in the stratosphere down to a cutoff pressure level 
  (refered to as ``Stratospheric'' hereafter). These profiles are not physically plausible (mainly due to the low homopause in Uranus, see next subsection), but have been considered for comparison 
  with \citet{Encrenaz2004} and \citet{Teanby2013}. Our results are described hereafter and are summarized in \tab{model_results}.
  
  The uniform distribution of \citet{Encrenaz2004} with a CO mole fraction of 2\dix{-8} overestimates the observed line core by a factor of 2.5-3. The observed line can be fitted with an 
  ``Uniform'' profile in which the CO mole fraction is 7.2\dix{-9} with \citet{Feuchtgruber2013}'s profile, or 9.3\dix{-9} with \citet{Orton2013a}'s profile.
  
  The line can be fitted equally well with a ``Stratospheric'' profile in which the CO is constant above the 100\,mbar level with a CO mole fraction of 7.1\dix{-9} with the thermal profile of 
  \citet{Feuchtgruber2013}, or 9.0\dix{-9} with \citet{Orton2013a}'s profile. For comparison with other papers (e.g. \citealt{Encrenaz2004,Cavalie2008a,Teanby2013}), we have set this transition level 
  to 100\,mbar, but our computations show this level could be located anywhere between $\sim$3 and 1000\,mbar, our results in terms of mole fraction would be affected by less than 10\%. If set above 
  the 3\,mbar level, then more CO would be needed.
  
\begin{figure}[!h]
     \begin{center}
     \includegraphics[width=9cm,keepaspectratio]{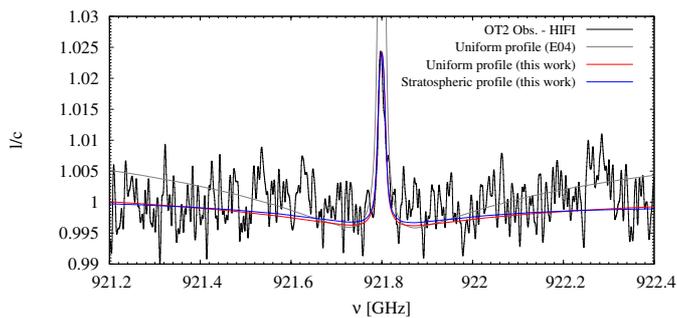}
     \end{center}
     \caption{Herschel-HIFI observation of the CO(8-7) line in Uranus on June 15, 2012, expressed in terms of line-to-continuum ratio ($l/c$, black line). This line can be modeled successfully 
     with either empirical models: (i) a ``Uniform'' profile with a constant mole fraction of 7.2\dix{-9} throughout the atmosphere (red line), and (ii) a ``Stratospheric'' profile with a constant mole fraction of 7.1\dix{-9} 
     above the 100\,mbar level and zero below it (blue line). The spectrum resulting from the \citet{Encrenaz2004} uniform source profile is also shown for comparison (grey line, labeled ``E04''). The synthetic lines have 
     been obtained with the thermal profile of \citet{Feuchtgruber2013}.}
     \label{empirical_models} 
  \end{figure}

  From these empirical models, it is not possible to favor an internal or an external origin for CO in the atmosphere of Uranus, because the models cannot be distinguished  (see \fig{empirical_models}).
  
  Fortunately, \citet{Teanby2013} recently published Herschel-SPIRE observations at CO line wavelengths. These observations are sensitive to the 10-2000\,mbar range, with a 
  contribution function peak at 200\,mbar (see their Fig.~2b), and they did not result in any detection. Those authors have set a stringent upper limit of 2.1\dix{-9} on the CO mole fraction in 
  their internal source model. This is $\sim$3-5 times lower than required by our observation. It is thus a clear indication that the HIFI line is caused by external CO. 

  \begin{table}
    \caption{Summary of the empirical and diffusion model results.}             
    \label{model_results}      
    \begin{center}          
    \begin{tabular}{cccc}
      \hline
      \multicolumn{4}{c}{Empirical model}\\
      \hline       
      Thermal profile & Uniform & \multicolumn{2}{c}{Stratospheric} \\
      \hline                    
      Feuchtgruber & 7.2\dix{-9}           & \multicolumn{2}{c}{7.1\dix{-9}} \\
      Orton               & 9.3\dix{-9}           & \multicolumn{2}{c}{9.0\dix{-9}} \\
      \hline  
      \hline
      \multicolumn{4}{c}{Diffusion model}\\
      \hline       
      Thermal profile                   & Internal source & \multicolumn{2}{c}{External source} \\
                                                    & \yco                     & \Flux{CO}   & $y_0$ \\
      \hline                    
      Feuchtgruber  & 1.9\dix{-8}          & 2.2\dix{5}  & 3.1\dix{-7} \\
      Orton                & 2.7\dix{-8}          & 2.7\dix{5}  & 3.9\dix{-7} \\
      \hline  
    \end{tabular}
    \end{center}
    \small{\underline{Note:} All results are mole fractions, except \Flux{CO} (in  cm$^{-2}\cdot$s$^{-1}$). The cut-off level in the ``Stratospheric'' empirical model is at 100\,mbar. 
    The internal source value of \yco~in the diffusion model enables fitting the CO line core amplitude, but the line is too broad and additional broad wings incompatible with the data 
    are generated.}
  \end{table}

  \subsection{Diffusion model}
  Uranus has the lowest homopause amongst the Giant Planets \citep{Orton1987,Moses2005}. It is located around the 1\,mbar level, where submillimeter observations generally probe. 
  Therefore, we have computed more realistic CO vertical profiles by accounting for diffusion processes to see how our previous results are impacted by the low homopause of 
  Uranus. Such modeling also shows how the various external sources can be parametrized.
  
  The vertical profile of CO primarily depends on the sources of CO, but it is also influenced by other oxygen sources. Indeed, O produced by H$_2$O photolysis reacts with CH$_3$ and 
  other hydrocarbons to produce CO \citep{Moses2000}. The magnitude of the H$_2$O flux is still quite uncertain, mostly due to limitations in the knowledge of the thermal structure and 
  the eddy mixing at the time of \citet{Feuchtgruber1997}'s observations. For the sake of simplicity, (photo-)chemical processes have been ignored. 
  
  We have adapted to Uranus the 1D time-dependent model of \citet{Dobrijevic2010,Dobrijevic2011} and removed all photochemical processes. \citet{Orton2013b} have constrained the 
  stratospheric $K_{zz}$ within [1000:1500]\,\Kunit~(vertically constant) with CH$_4$ and C$_2$H$_6$ Spitzer observations. We have taken their best fit value (1200\,\Kunit) in our 
  model. Three sources of CO, representing simple cases, have been tested: (i) an internal source, (ii) a steady flux of micrometeorites (IDP), and (iii) a single large comet impact\footnote{This 
  does not exclude a combination of internal and external sources, or any intermediate situation between a continuum of micrometeoritic impacts and a single impact event.}. The three sources 
  tested are controlled by a few parameters: (i) the tropospheric CO mole fraction \yco~for an internal source, (ii) the flux \Flux{CO} at the upper boundary of the model atmosphere for a steady 
  source, and (iii) the equivalent mole fraction of CO $y_0$ deposited by a comet and averaged over the planet. We have assumed that all the CO was deposited at levels higher than 0.1\,mbar 
  in analogy to the SL9 impacts \citep{Lellouch1995,Moreno2003} and that the impact time $\Delta t$$\sim$300\,years as it roughly corresponds to the diffusion time down to 1\,mbar in Uranus 
  in our model, but other combinations of deposition time and level are possible. To infer the mass and diameter of the comet, we have assumed the comet density was 0.5\,g$\cdot$cm$^{-3}$ 
  \citep{Weissman2004,Davidsson2007} and that the comet yielded 50\% CO at impact \citep{Lellouch1997}.
  
\begin{figure*}[!h]
   \begin{center}
   \includegraphics[width=18cm,keepaspectratio]{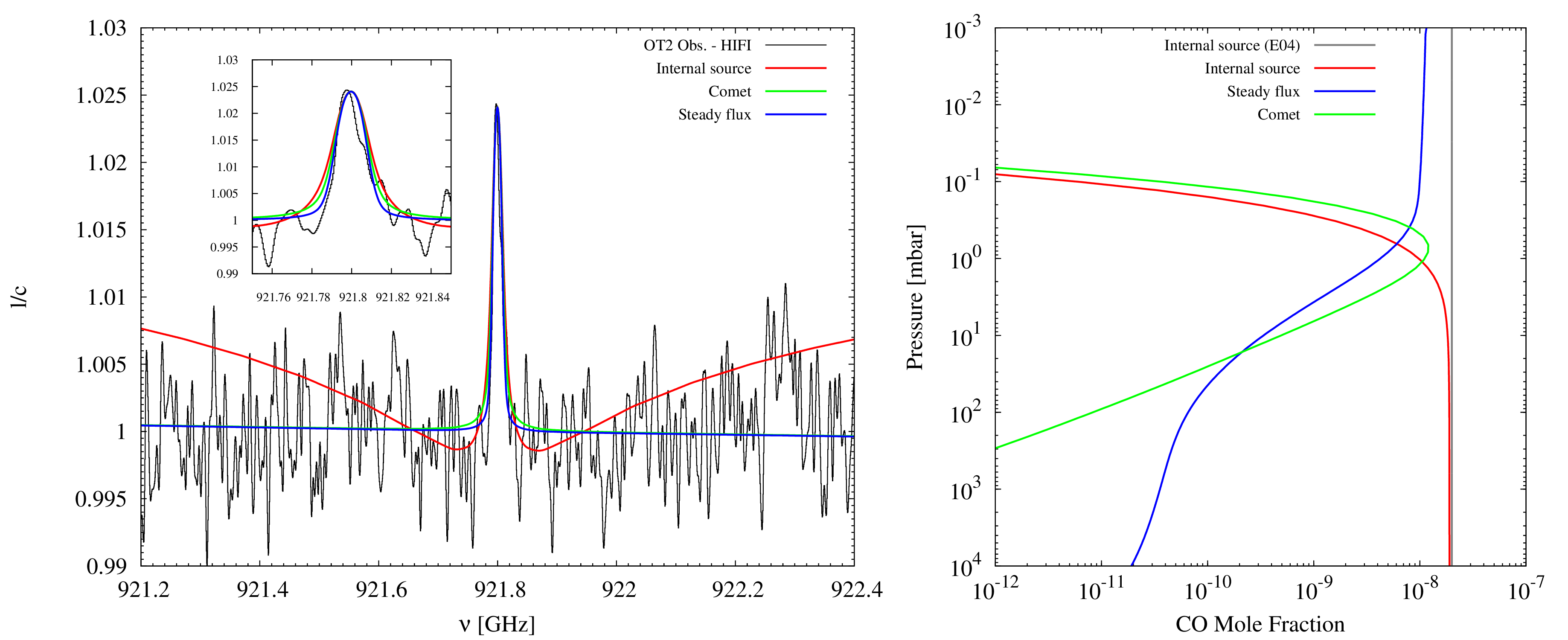}
   \end{center}
   \caption{Left: Herschel-HIFI observation of the CO(8-7) line in Uranus on June 15, 2012, expressed in terms of line-to-continuum ratio ($l/c$, black line). For each source, the models that best fit the emission 
   core are displayed: an internal source yielding a mole fraction of 1.9\dix{-8} in the upper troposphere (red line), a steady external flux (due to IDP or a local source) of 2.2\dix{5}\,\Fluxunit~(blue line), and a comet 
   with a diameter of 640\,m depositing 3.4\dix{13}\,g of CO above the 0.1\,mbar level $\sim$300\,years ago (green line). These models have been computed with the thermal profile of \citet{Feuchtgruber2013}. 
   The internal source model overestimates the line core width and produces a broad absorption that is not observed in the data. The external source models can barely be differentiated. Right: Vertical profiles 
   associated to the spectra.}
   \label{CO_HIFI_2012} 
\end{figure*}

  The vertical profiles and resulting spectra corresponding to the three sources, as obtained with the thermal profile of \citet{Feuchtgruber2013}, are displayed in \fig{CO_HIFI_2012}. The 
  best fits to the spectrum are obtained for external source models. Despite resulting in different vertical profiles, a steady flux \Flux{CO}$=$2.2\dix{5}\,\Fluxunit~and a 640\,m diameter comet 
  depositing 3.5\dix{13}\,g of CO ($y_0$$=$3.1\dix{-7}) result in lines that cannot be distinguished from one another with our observations. Such impact at Uranus occurs every $\sim$500~years 
  with a factor of 6 uncertainty \citep{Zahnle2003}. Such timescales are fully compatible with our assumption on $\Delta t$. With the thermal profile of \citet{Orton2013a}, we obtain slightly higher 
  values because of lower stratospheric temperatures: \Flux{CO}$=$2.7\dix{5}\,\Fluxunit~and $y_0$$=$3.9\dix{-7} (i.e. a 700\,m diameter comet). All fit parameters are listed in \tab{model_results}. 
  These values remain to be confirmed by more rigorous (photochemical) modeling and higher S/N data. 
  
  The amplitude of the CO core emission is reproduced with an internal source model in which \yco$=$1.9\dix{-8} (see \fig{CO_HIFI_2012}). With \citet{Orton2013a}'s thermal profile, \yco$=$2.7\dix{-8}. 
  We note that $\sim$3 times more tropospheric CO is needed in this model, compared to the ``Uniform'' empirical model value derived in Sect.~\ref{empirical}. This is due to the fact that the observed emission line probes 
  the mbar level, i.e. where the CO vertical profile sharply decreases because of the low homopause in the atmosphere of Uranus. As a result, a stronger internal source is required to reach a sufficient level of 
  abundance of CO around the mbar level. The main outcome of this model is that it now overestimates the line core width and results in an additional broad absorption, because CO is much more abundant in 
  the lower stratosphere than in the external source models (by already 2 orders of magnitude at 10\,mbar). The absence of such a broad CO absorption in the data cannot be caused by our sinusoidal ripple 
  removal procedure, because we have removed sine waves of much shorter period than the total width of such broad absorption wings. Because the width of the line core is not fitted and because there is no 
  broad absorption in the spectrum, and because the derived \yco values are an order of magnitude larger than the upper limit set by Herschel-SPIRE observations of this region of the atmosphere 
  \citep{Teanby2013}, the internal source model can be ruled out. Thus, as long as there is no significant photochemical source of CO in the stratosphere, the HIFI line is caused by external CO.

  \subsection{An upper limit on the deep O/H ratio in Uranus} 
  Thermochemistry in the deep interior of Uranus links the CO abundance to the one of H$_2$O and thus to the internal O/H ratio \citep{Fegley1988,Lodders1994} with the following net thermochemical 
  equilibrium reaction
  \begin{equation*}
    \mathrm{H}_2\mathrm{O}+\mathrm{CH}_4=\mathrm{CO}+3\mathrm{H}_2.
  \end{equation*}
  The upper tropospheric mole fraction of CO is fixed at the level where the thermochemical equilibrium is quenched by vertical diffusion.
  
  The upper limit of \citet{Teanby2013} on the internal source (\yco$=$2.1\dix{-9}) can be further used to try and constrain the deep atmospheric O/H ratio in Uranus. Their observations probe between 10 
  and 2000\,mbar, i.e. well below the homopause level (see \fig{CO_HIFI_2012} right). As a consequence, this upper limit is valid even if the authors have not accounted for the low homopause of Uranus.
  
\begin{figure*}[!h]
     \begin{center}
     \includegraphics[width=12cm,keepaspectratio]{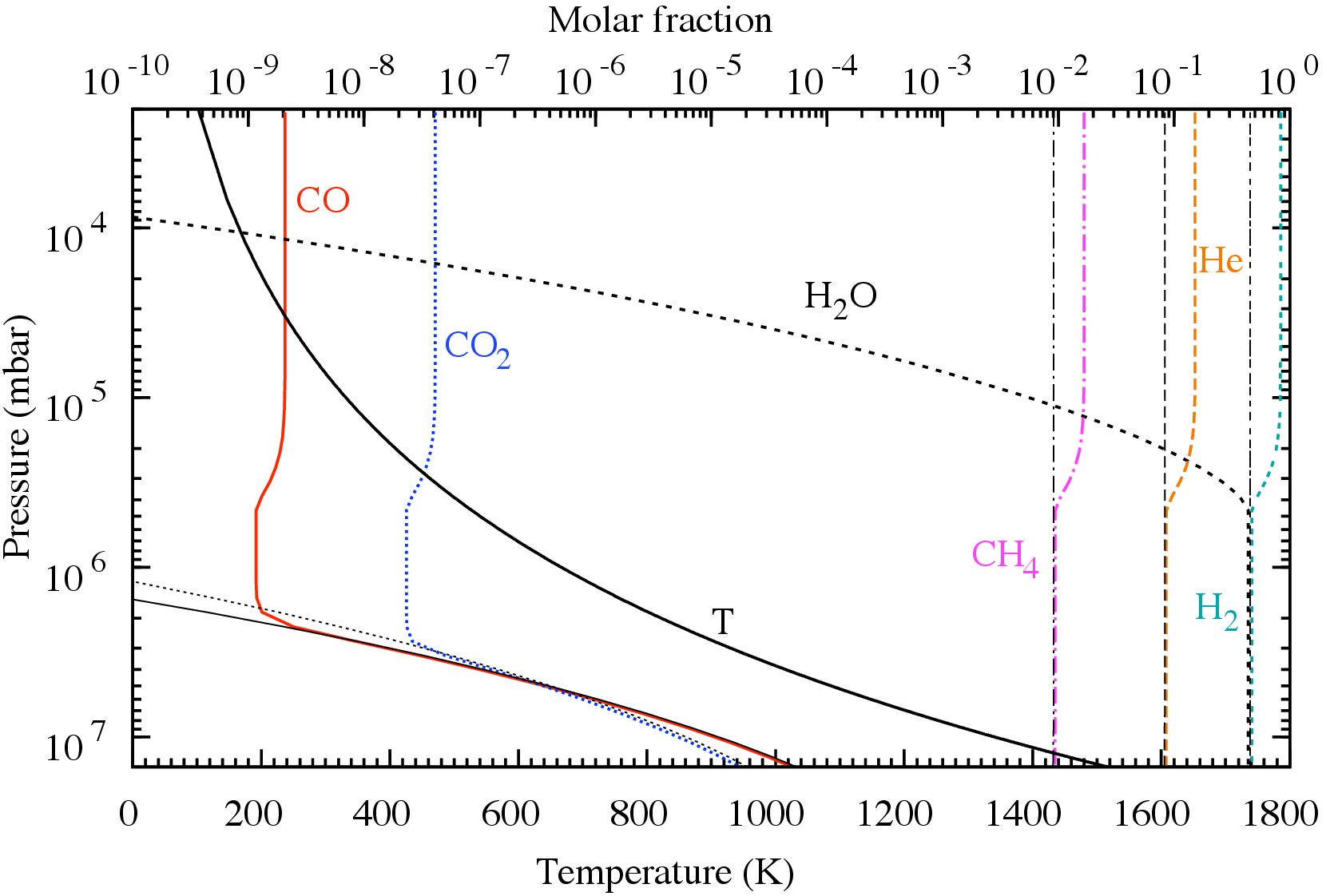}
     \end{center}
     \caption{Molar fraction profiles in the troposphere of Uranus obtained with \citet{Venot2012}'s model, targeting the 2.1\dix{-9} upper limit on the upper tropospheric CO mole fraction obtained by \citet{Teanby2013}. 
     The temperature profile in the troposphere is shown in black solid line. Thermochemical equilibrium profiles are plotted in black with the same layout as their corresponding species. CO and CO$_2$ 
     are quenched around 2-3\dix{6}\,mbar. H$_2$O departs from thermochemical equilibrium because of condensation and causes the increase of other species mole fractions (the sum of all 
     mole fractions is normalized to unity at all levels). The model parameters are: O/H$=$501$\Sun$, C/H$=$18$\Sun$, and $K_{zz}$$=$$10^8$\,\Kunit.}
     \label{Thermo} 
  \end{figure*}

  We have adapted to Uranus the thermochemical model developed by \citet{Venot2012} to constrain the O/H ratio. This model accounts for C, N and O species. We have extended our thermal profile to high 
  pressures following the dry adiabat (the \citealt{Feuchtgruber2013} and \citealt{Orton2013a} are similar in the upper troposphere and thus give similar deep tropospheric profiles) and we have constrained 
  the O/H and C/H ratios by fitting the following upper tropospheric mole fractions with errors lower than 4\%: 0.152 for He \citep{Conrath1987}, 0.016 for CH$_4$ \citep{Baines1995,Sromovsky2008}, 
  and the 2.1\dix{-9} upper limit for CO. The level at which CO is quenched depends not only on the temperature profile and the deep O/H ratio, but also on the deep $K_{zz}$. Assuming Uranus' interior is 
  convective, we can estimate $K_{zz}$ from the planet's internal heat flux \citep{Stone1976}. Following \citet{Pearl1990}, $K_{zz}$$\sim$10$^8$\Kunit, within one order of magnitude \citep{Lodders1994}. 
  The resulting tropospheric vertical profiles for this nominal model are shown in \fig{Thermo}. The elemental ratios in this model are 501 times solar for O/H and 18 times solar for C/H (with 
  solar abundances, $\Sun$ hereafter, taken from \citealt{Asplund2009}). The N species have no significant impact on the C and O species. We have also computed the elemental ratios in a series of additional 
  models to evaluate the influence of parameters like $K_{zz}$ and the upper tropospheric CH$_4$ mole fraction on the O/H ratio. The results are displayed in \tab{elemental_ratios}. We find that the deep O/H 
  is lower than $\sim$500$\Sun$ (nominal model), but could be even below 340$\Sun$ to be in agreement with the CO tropospheric upper limit in all cases. On Neptune, \citet{Luszcz-Cook2013} find that ``an upwelled 
  CO mole fraction of 0.1 ppm implies a global O/H enrichment of at least 400, and likely more than 650, times the protosolar value''.

  \begin{table}
    \caption{Summary of the thermochemical model results. The values have been obtained so as to reach the 2.1\dix{-9} upper limit of \citet{Teanby2013} for the CO upper tropospheric mole fraction.}             
    \label{elemental_ratios}      
    \begin{center}          
    \begin{tabular}{cccccc}
      \hline       
      Model & $K_{zz}$ & $y_{\mathrm{CH}_4}$ & C/H  & \yco  & O/H \\
       & \Kunit & \dix{-2} & $\times\Sun^{(1)}$  &  \dix{-9} & $\times\Sun^{(2)}$ \\
      \hline                    
      Nominal                   & 10$^8$   &1.6$^{(3)}$ & 18 & 2.1    & 501 \\
      CH$_4$-rich           & 10$^8$   & 3.2$^{(4)}$ & 40 & 2.1    & 417         \\
      low $K_{zz}$           & 10$^7$   & 1.6               & 13 & 2.1    & 631         \\
      high $K_{zz}$         & 10$^9$   & 1.6               & 23  & 2.1    & 339          \\
      \hline  
    \end{tabular}
    \end{center}
    \small{$^{(1)}$ Solar C/H volume ratio: 2.69\dix{-4} \citep{Asplund2009}\\
    $^{(2)}$ Solar O/H volume ratio: 4.90\dix{-4} \citep{Asplund2009}\\
    $^{(3)}$ \citet{Baines1995} and \citet{Sromovsky2008}\\
    $^{(4)}$ \citet{Fry2013}, \citet{Sromovsky2011}, and \citet{Karkoschka2009}}
  \end{table}
  
\section{Discussion and conclusion \label{Discussion}}
We have detected the CO(8-7) line at 921.800\,GHz in Uranus with Herschel and we have constrained its possible sources. 

Herschel-HIFI (this work) and Herschel-SPIRE \citep{Teanby2013} results show that the average CO mole fraction is decreasing from the stratosphere to the troposphere. This suggests the deep interior is 
not the source of the observed CO. Our diffusion model calculations confirms that the internal source hypothesis is not valid and shows that Uranus has an external source of CO as long as there is not a 
significant photochemical source of CO in the stratosphere. The data can be successfully fitted by an empirical model in which CO has a mole fraction of 7.1-9.0\dix{-9} above the 100\,mbar level 
(value depending on the chosen thermal profile). There is a contradiction between this model mole fraction values and the mole fraction reported by \citet{Encrenaz2004} (3\dix{-8} in their external source 
model). Regarding this apparent discrepancy, we note that modelling LTE emission from CO rotational lines is much simpler than inferring an abundance from non-LTE fluorescence (e.g. 
\citealt{Lopez-Valverde2005}). At any rate, a reanalysis of the \citet{Encrenaz2004} data in the light of CO distributions proposed in this paper should be performed.

Comet and steady source models, in which diffusion processes are accounted for, give very similar fit to the data. These results should be confirmed with more elaborate models, i.e. photochemical 
models, and more sensitive observations. Oxygen photochemistry computations, taking into account nearly concomitant measurements of the thermal profile \citep{Feuchtgruber2013,Orton2013a}, of the 
influx of H$_2$O (Jarchow et al., in prep.), and of the influx of CO$_2$ \citep{Orton2013b} would enable us to draw better constraints on the external source of oxygen. It would certainly reduce the external 
flux of CO or the mass of the impacting comet we have obtained from a simple diffusion model, because the chemical conversion of H$_2$O into CO would already provide a significant part of the observed 
stratospheric column of CO. 

The internal source upper limit derived by \citet{Teanby2013} (\yco$=$2.1\dix{-9}), which also goes in contradiction with the detection level of \citet{Encrenaz2004} (2\dix{-8} in their internal source model), 
was used to derive an upper limit on the deep O/H ratio of Uranus. Our thermochemical simulations show that the deep O/H ratio is lower than 500$\Sun$ to provide a \yco~value lower than 2.1\dix{-9}. 
A dedicated probe, as in the mission concepts proposed by \citet{Arridge2013} and \citet{Mousis2013} in response to the ESA 2013 Call for White Papers for the Definition of the L2 and L3 Missions in the 
ESA Science Programme, or radio observations might be the only way to measure the deep O/H ratio in Uranus.


\begin{acknowledgements}
  T. Cavali\'e acknowledges funding from the Centre National d'\'Etudes Spatiales (CNES). T. Cavali\'e and F. Selsis acknowledge support from the European 
  Research Council (Starting Grant 209622: E$_3$ARTHs). O. Venot acknowledges support from the KU Leuven IDO project 
  IDO/10/2013 and from the FWO Postdoctoral Fellowship programme. G. Orton acknowledges funding from the National 
  Aeronautics and Space Administration to the Jet Propulsion Laboratory, California Institute of Technology. L.~N. Fletcher was 
  supported by a Royal Society Research Fellowship at the university of Oxford. The authors thank M. Hofstadter for his constructive 
  review. HIFI has been designed and built by a consortium of institutes and university departments from across Europe, Canada 
  and the United States under the leadership of SRON Netherlands Institute for Space Research, Groningen, The Netherlands and 
  with major contributions from Germany, France and the US. Consortium members are: Canada: CSA, U.Waterloo; France: CESR, 
  LAB, LERMA, IRAM; Germany: KOSMA, MPIfR, MPS; Ireland, NUI Maynooth; Italy: ASI, IFSI-INAF, Osservatorio Astrofisico di 
  Arcetri-INAF; Netherlands: SRON, TUD; Poland: CAMK, CBK; Spain: Observatorio Astron\'omico Nacional (IGN), Centro de 
  Astrobiolog\'ia (CSIC-INTA). Sweden: Chalmers University of Technology - MC2, RSS \& GARD; Onsala Space Observatory; 
  Swedish National Space Board, Stockholm University - Stockholm Observatory; Switzerland: ETH Zurich, FHNW; USA: Caltech, 
  JPL, NHSC. The James Clerk Maxwell Telescope is operated by the Joint Astronomy Centre on behalf of the Science and 
  Technology Facilities Council of the United Kingdom, the National Research Council of Canada, and the Netherlands Organisation 
  for Scientific Research.
\end{acknowledgements}


\end{document}